\documentclass[aps,reprint,showkeys]{revtex4-1}
\usepackage{graphicx}
\usepackage{amssymb}
\usepackage{amsmath}
\usepackage{bm}
\usepackage{color}
\usepackage{hyperref}
\usepackage[german, english]{babel}
\usepackage[section]{placeins}
\usepackage{caption}

\begin{document}
\setcounter{page}{1}
\title[]{Inner-most stable circular orbit in Kerr-MOG black hole}
\author{Hyun-Chul \surname{Lee}}
\author{Yong-Jin \surname{Han}}
\email{yjhan7@sch.ac.kr}
\thanks{Fax: +82-41-530-1229}
\affiliation{Department of Physics, Soonchunhyang University, Asan
646}

\date{\today}

\begin{abstract}
We study the inner-most stable circular orbit(ISCO) of Kerr black hole in MOdified gravity(Kerr-MOG black hole) which is one of the exact solution of the field equation of modified gravity in strong gravity regime. Kerr-MOG black hole is constructed commonly known kerr metric in Boyer-Lindquist coordinate by adding repulsive parameter like Yukawa force which explained in quantum gravity. We explore what is some differences of the ISCOs of Kerr-MOG black hole and kerr black hole.

\end{abstract}

\pacs{84.40.Ik, 84.40.Fe}

\keywords{Modified Gravity, Black hole, General relativity, Inner-most stable circular orbit}

\maketitle

\section{INTRODUCTION}
General relativity(GR) is well-established theory by predicting many cosmological phenomena in solar system and universe. Specially, it predicts a black hole which is one of astronomically the most surprising phenomena. However, interior of black hole and some astronomical observations are not explained by GR.
In particular, a strange phenomenon of galaxy rotation curves found by the study of Zwicky, Rubin and her collaborators is the example.\cite{R1,R2,R3,R4}. Zwicky, first theoretically and formally postulated unknown matter, called dark matter and Rubin found that Einstein's theory can not explain the rotation curve of a spiral galaxy. This is an important clue for the existence of dark matter. However, dark matter has not discovered by experimental observation so far because it is not interacting with visible matter. One needed explainable theory to this phenomena without invoking the concept of dark matter.
First of all, Modified Newtonian Dynamics(MOND) is proposed by Milgorom by altering “inverse-square” law of gravity in scalar field of Newtonian gravity\cite{R5}. This fits on some observations and weak-field approximation like solar system, but MOND has limit that explain the phenomena like a large galactocentric distance of $\sim 200 kpc$ and globular cluster \cite{R6,R7}.

Scalar-tensor-vector gravity(STVG) theory was proposed by Moffat\cite{R8}. STVG referred as the Modified gravity theory(MOG) evolved to explain the discordance between the general relativity at large scale and many astronomical observations such as dynamics of galaxies, galaxy rotation curves, bullet cluster, the amount of luminous matter of these galaxies and accelerating universe\cite{R8,R9,R10}. MOG theory is the field theory which is added, scalar action, vector action and matter action, thereby modifying Einstein-Hilbert action.
Moffat also obtained the metric of Kerr modified gravity (Kerr-MOG) of black hole by combining the  MOG theory with Kerr geometry\cite{R11}. This theory is the alternative theory in strong gravity and exact solution for rotating black hole. He also theoretically showed observational methods for black hole shadow of Kerr-MOG\cite{R12}. The research for shadow of black hole at our galactic center is in progress by Event Horizon Telescope(EHT)\cite{R13}. EHT can be expected to observe shadow of a black hole by particles or photons around the black hole accretion disk.
One can not observed directly a black hole, thus someone must study properties near black hole. Most of all, movement of a particle or photon by strongly curved spacetime are useful subjects. In particular, circular motions of timelike or lightlike geodesics are important because it gives us information about geometric framework around non-rotating or rotating black holes.
Among them, there is the minimum radius of trajectory that is able to maintain stable circular orbit and do not enter into event horizon of black hole called innermost stable circular orbit(ISCO)\cite{R14}.
In the accretion disk theory\cite{R15}, ISCO is regarded as one of rotating black hole's important features such as event horizon, erosphere, etc. It is believed to be the inner edge of an accretion disc orbiting black hole. Iron line profiles for black holes is determined the photons that are emitted near the ISCO\cite{R16}.
The radius of ISCO is $6M$ in the case of schwarzschild black hole while radius of ISCO for Kerr black hole is depend on spin $a$. As an example, extreme rotating black hole, that is intrinsic angular momentum of kerr black hole is one $(a=1)$, have $M$ in co-rotating and $9M$ in counter-rotating where $G_{N}=c=1$ in case of Schwarzschild and Kerr metric\cite{R17}.
 ISCO assist the study which find shape of thin disk and Penrose process at erosphere as well as properties in the vicinity of the black hole.

 In this paper, we investigate ISCO of Kerr-MOG black hole that is not yet obtained. We obtain the expression of energy $E$ and angular momentum $L_{z}$ of test particle in circular orbit for Kerr-MOG metric, using a method of Ref.\cite{R18,R19}. We also numerically calculate ISCO of Kerr-MOG and analyze the result.

\section{FIELD EQUATION OF MOG AND KERR-MOG METRIC}

The field equation of MOG are given by \cite{R8,R11}
\begin{equation}
\begin{split}
G_{\mu\nu} = -8\pi GT_{\phi\mu\nu}
\end{split}
\end{equation}

where $c=1$ and postulated that $G=G_{N}(1+\alpha)$, $\partial_{\nu}G=0$ where $G_{N}$ is Newton's gravitational constant and $G$ is deformed as parameter $\alpha$. Furthermore, the canonical energy-momentum tensor of matter $T_{M\mu\nu}=0$ and the energy-momentum tensor for the vector field $\phi_{\mu}$ are induced by

\begin{equation}
\begin{split}
T_{\phi\mu\nu} = -\frac{1}{4\pi}\Big(B_{\mu}^{\;\; \sigma}B_{\nu\sigma} - \frac{1}{4}g_{\mu\nu}B^{\sigma\beta}B_{\sigma\beta}\Big)
\end{split}
\end{equation}

where $B_{\mu\nu} = \partial_{\mu}\phi_{\nu} - \partial_{\nu}\phi_{\mu}$ and  $\phi_{\mu}$ are vector field with the source charge $Q = \sqrt{\alpha G_{N}}M$. Moreover, the vacuum field equations are

\begin{equation}
\begin{split}
&\nabla_{\nu}B^{\mu\nu} = \frac{1}{\sqrt{-g}}\partial{_\nu}(\sqrt{-g}B^{\mu\nu}) = 0,
\\\\
&\nabla_{\sigma}B_{\mu\nu} + \nabla_{\mu}B_{\nu\sigma} + \nabla_{\nu}B_{\sigma\mu} = 0
\end{split}
\end{equation}
where $\nabla_{\nu}$ is in regard to the metric tensor $g_{\mu\nu}$.

The Kerr-MOG black hole is static and axisymmetric solution of gravitational field equation. Kerr-MOG metric in Boyer-Lindquist coordinates is

\begin{equation}
\begin{split}
& ds^{2} = \frac{\Delta}{\rho^{2}}(dt - a\sin^{2}{\theta}d\phi)^{2}\ \\\\
& -\frac{\sin^{2}\theta}{\rho^{2}}[(r^{2}+a^{2})d\phi - adt]^{2} - \frac{\rho^{2}}{\Delta}dr^{2} - \rho^{2}d\theta^{2}
\end{split}
\end{equation}
where
\begin{equation}
\begin{split}
&\Delta = r^{2}-2G_{N}(1+\alpha)Mr+a^{2}+M^{2}G_{N}^{2}\alpha(1+\alpha),\\
& \rho^{2}=r^2+a^{2}\cos^{2}{\theta}.
\end{split}
\end{equation}


\section{TOTAL ENERGY OF Kerr-MOG METRIC IN CIRCULAR ORBIT}

To usually find circular orbit in any metric, one considers effective potential for motion of massive particle or photon\cite{R14,R15,R16,R17,R18,R19,R20,R21,R22,R23,R24,R25,R26}. We approach the problem by solving ISCO by well-developed the method differently than Moffat did\cite{R11,R18,R19}.
We investigate behavior of a test particle in equatorial plane ($\theta=\pi/2$) and thus covariant components of Kerr-MOG metric are

\begin{equation}
\begin{split}
& g_{tt}= \frac{\Delta-a^{2}}{r^{2}},\ \ g_{t\phi}= \frac{a}{r^{2}}{\Big[(r^{2}+a^{2})-\Delta\Big]},\  \ g_{rr}= -\frac{r^{2}}{\Delta},\\\\
& g_{\theta\theta}=-r^{2},\ \ g_{\phi\phi}= -\frac{1}{r^2} \Big[(r^{2}+a^{2})^{2}-a^{2}\Delta \Big]. \\
\end{split}
\end{equation}

Because of the property of stationary and axis-symmetric for Kerr-MOG metric , we consider that two conserved quantities are the specific energy $E$ and specific angular momentum $L_{z}$ for constant motion of the test particle.
We use the Lagrangian and the Euler-Lagrange equations for solving geodesic equation.

\begin{equation}
L = \frac{1}{2}g_{\mu\nu} \dot{x}^{\mu}\dot{x}^{\nu}.
\end{equation}

In this way, we derive Equation for $t$ and $\phi$ as well as the radial equation.

\begin{equation}
\begin{split}
&\dot{t} = \frac{-Eg_{\phi\phi} - L_{z}g_{t\phi}}{g_{t\phi}^{2} - g_{tt}g_{\phi\phi}}\\\\
&\dot{\phi} = \frac{Eg_{t\phi} + L_{z}g_{tt}}{g_{t\phi}^{2} - g_{tt}g_{\phi\phi}}\\\\
&\dot{r}^{2} = \frac{1}{r^2} \Bigg[E^{2}r^{2} - (L_{z}^{2}-a^{2}E^{2})+(\frac{2M_{\alpha}}{r} -\frac{\beta^{2}}{r^{2}})\times\\
&(L_{z}-aE)^{2} - \Delta\mu \Bigg]
\end{split}
\end{equation}
where
\begin{equation}
\begin{split}
M_{\alpha} = G_{N}(1+\alpha)M ,\  \ \beta^{2} = M^2G_{N}^{2}\alpha(1+\alpha).
\end{split}
\end{equation}

For convenient calculation, we temporarily utilize above values and we study timelike geodesics $(\mu=1)$ of the circular orbit of the test particle. We will do reciprocal radius $u(=1/r)$ and $x=L_{z}-aE$ as independent variable and we consider $\dot{r}=0$ and $\partial\dot{r}/\partial r=0$. In next section, we should be explain in detailed. Radial equations are

\begin{equation}
\begin{split}
E^2-(x^{2}+2aEx)u^{2} +(2M_{\alpha}-\beta^{2}u)x^{2}u^{3}\\
-(1+a^{2}u^{2}-2M_{\alpha}u+\beta^2u^{2})=0
\end{split}
\end{equation}

and

\begin{equation}
\begin{split}
 -(x^{2}+2aEx)u^{2} +3M_{\alpha}x^{2}u^{3}-2\beta^{2}x^{2}u^{4}\\
 -(a^{2}u^{2}-M_{\alpha}u+\beta^2u^{2})=0.
\end{split}
\end{equation}

\captionsetup[figure]{labelfont={bf},labelformat={default},labelsep=period,name={Fig.}}
\captionsetup{justification=raggedright,singlelinecheck=false}
\begin{figure}[t]
  \begin{center}
  \includegraphics[width=\columnwidth]{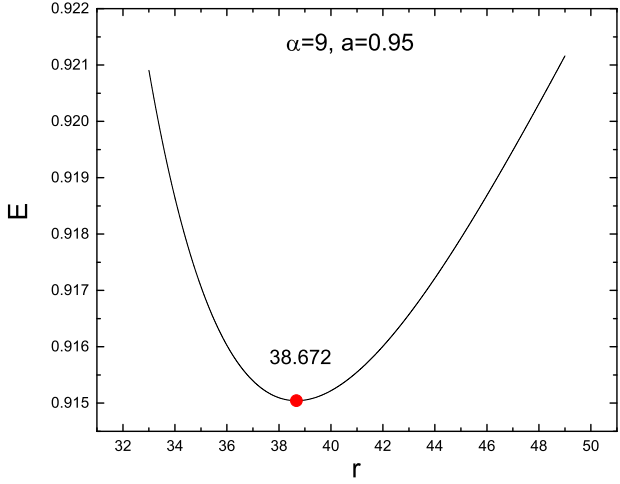}
  \caption{{\bf Energy $E$ is a function of $r$.} For $\alpha=9$, $a=0.95$, there is minimum point at ISCO $r=38.672$.}
   \label{fig.1}
\end{center}
\end{figure}

Subtracting Eq. (11) from Eq. (10), we take

\begin{equation}
\begin{split}
E^{2} = 1-M_{\alpha}u + M_{\alpha}x^{2}u^{3}-\beta^{2}x^{2}u^{4}
\end{split}
\end{equation}

and from Eq.(11), we obtain

\begin{equation}
\begin{split}
2aExu = x^2\Big[3M_{\alpha}u - 2\beta^{2}u^{2} - 1 \Big]u -(a^{2}u - M_{\alpha}) - \beta^{2}u.
\end{split}
\end{equation}

We also can derive quadratic equation for $x$ from combining Eq. (12) with Eq. (13) and eliminating $E$

\begin{equation}
\begin{split}
& x^4u^2\Big[(3M_{\alpha}u-1)^{2} - 4a^{2}M_{\alpha}u^{3} + 4\beta^{4}u^{4} \\
&-4\beta^{2}u^{2}(3M_{\alpha}u-1) + 4a^{2}\beta^{2}u^{4} \Big] +(a^{2}u - M_{\alpha})^{2} + \beta^{4}u^{2}\\
&+ 2\beta^{2}u(a^{2}u-M_{\alpha})-2x^{2}u\Big[(3M_{\alpha}u-1)(a^{2}u-M_{\alpha})\\
&+ 2a^{2}u - 2Ma^{2}u^{2} - 2\beta^2u^{2}(a^{2}u - M_{\alpha})+\beta^{2}u(3M_{\alpha}u-1)\\
&- 2\beta^{4}u^{3}\Big] = 0.
\end{split}
\end{equation}

The discriminant "$\frac{1}{4}(b^{2}-4ac)$" of this equation is

\begin{equation}
\begin{split}
4a^{2}u^{3}(M_{\alpha} - \beta^{2}u)\Delta_{u}^{2}
\end{split}
\end{equation}
where
\begin{equation}
\begin{split}
\Delta_{u} = a^{2}u^{2} - 2M_{\alpha}u + 1 +\beta^{2}u^{2}.
\end{split}
\end{equation}
We take
\begin{equation}
\begin{split}
x^{2}u^{2} =  \frac{Q_{\pm}\Delta_{u}-Q_{+}Q_{-}}{Q_{+}Q_{-}} = \frac{1}{Q_{\mp}}(\Delta_{u} - Q_{\mp})
\end{split}
\end{equation}
where
\begin{equation}
\begin{split}
& Q_{+}Q_{-} = (3M_{\alpha}u - 2\beta^2u^{2}-1)^{2} - 4a^{2}u^{3}(M_{\alpha} - \beta^{2}u)\\\\
& Q_{\pm} = 1 + 2\beta^{2}u^{2} - 3M_{\alpha}u \pm 2a\sqrt{(M_{\alpha} - \beta^{2}u)u^{3}}
\end{split}
\end{equation}
and
\begin{equation}
\begin{split}
\Delta_{u} - Q_{\mp} = u\Big[a\sqrt{u} \pm \sqrt{(M_{\alpha} - \beta^{2}u)}\Big]^{2}.
\end{split}
\end{equation}

\begin{figure}[t]
  \begin{center}
  \includegraphics[width=\columnwidth]{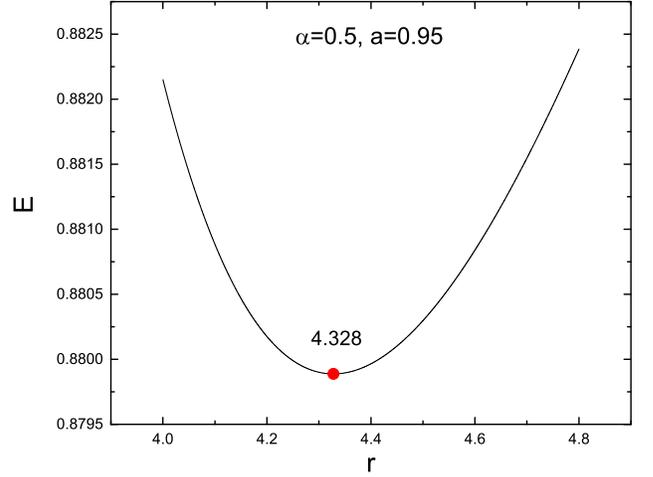}
  \caption{{\bf Energy $E$ is a function of $r$.} For $\alpha=0.5$, $a=0.95$, there is minimum point of effective potential equal to ISCO at $r=4.328$. }
   \label{fig.2}
\end{center}
\end{figure}

Therefore, we can find  the solution of  quadratic equation for $x$

\begin{equation}
\begin{split}
x = - \frac{a\sqrt{u} \pm \sqrt{(M_{\alpha} - \beta^{2}u)}}{\sqrt{uQ_{\mp}}}.
\end{split}
\end{equation}\\

The Eq. (20) shows that the upper sign in the foregoing equation refers to the counter-rotating orbit; On the other hand, the lower sign refers to the co-rotating orbit. We will use to the the lower sign to find ISCO of Kerr-MOG but now this convention leave a this section.
We can derive the equation of energy $E$ by substituting Eq. (20) in Eq. (12), changing Eq. (9)

\begin{equation}
\begin{split}
E = & \frac{1}{\sqrt{Q_{\mp}}}\Bigg[1 - 2G_{N}(1+\alpha)Mu + M^{2}G_{N}^2\alpha(1+\alpha)u^{2}\\
& \mp a\sqrt{\Big[G_{N}(1+\alpha)M - M^{2}G_{N}^{2}\alpha(1+\alpha)u\Big]u^{3}}\Bigg].
\end{split}
\end{equation}

According to $x = L_{z}-aE$, we find

\begin{equation}
\begin{split}
L_{z} = &\mp\frac{\sqrt{G_{N}M(1+\alpha)-M^{2}G_{N}^2\alpha(1+\alpha)u}}{\sqrt{uQ_{\mp}}}\Bigg [a^{2}u^{2} + 1 \\
& \pm\frac{a\sqrt{u^{3}}}{\sqrt{G_{N}M(1+\alpha)-M^{2}G_{N}^2\alpha(1+\alpha)u}} \Big[2G_{N}M(1 \\
&+\alpha)-M^{2}G_{N}^2\alpha(1+\alpha)u \Big ] \Bigg ]
\end{split}
\end{equation}

\begin{figure}[t]
  \begin{center}
  \includegraphics[width=\columnwidth]{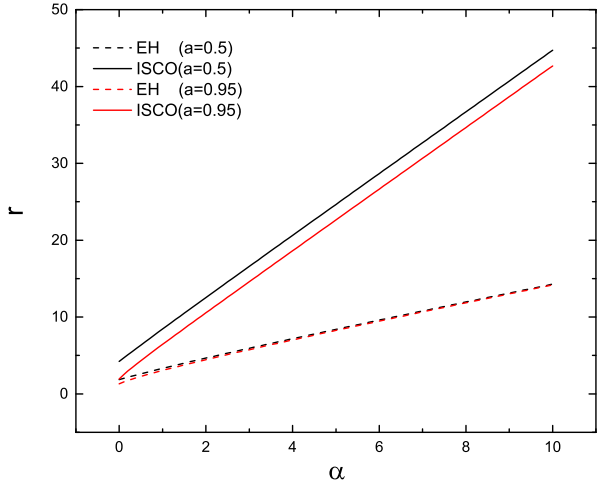}
  \caption{{\bf Event horizon(EH) and ISCO.} Distance between ISCO and event horizon of Kerr-MOG black hole increases as parameter $\alpha$ increases}
  \label{fig.3}
\end{center}
\end{figure}

we easily see Kerr metric when $\alpha$ is zero,

\begin{equation}
\begin{split}
E = \frac{1}{\sqrt{Q_{\mp}}}\Big[1 - 2G_{N}Mu \mp a\sqrt{G_{N}Mu^{3}}\Big].
\end{split}
\end{equation}
and
\begin{equation}
\begin{split}
L_{z} = \mp\frac{\sqrt{G_{N}M}}{\sqrt{uQ_{\mp}}}\Big[a^{2}u^{2} + 1 \pm 2a\sqrt{G_{N}Mu^{3}}\Big].
\end{split}
\end{equation}

\section{INNER-MOST STABLE CIRCULAR ORBIT(ISCO) OF KERR-MOG BLACK HOLE}

We use to the method of Ref.\cite{R27}, to numerically derive radius of ISCO. Thus, we obtain numerical values of ISCOs of Kerr-MOG black holes from Eg.(21). Differentiating $E$ for radius $r$ is equal to zero, that is,

\begin{equation}
\begin{split}
\frac{dE}{dr} = 0.
\end{split}
\end{equation}
By condition (25), the radius of ISCO$(r_{ISCO})$ is obtained at minimum value of $E$.

\begin{figure}[t]
  \begin{center}
  \includegraphics[width=\columnwidth]{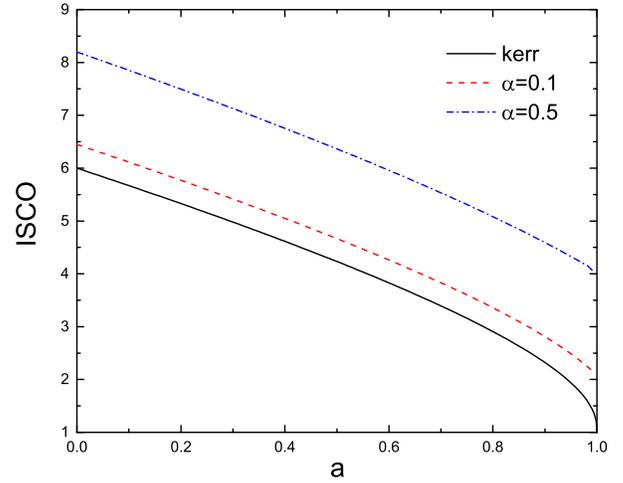}
  \caption{{\bf The case of kerr black hole is comparison with kerr-MOG black hole.} ISCO is plotted on some intrinsic angular momentum $a$ and parameter $\alpha$.}
   \label{fig.4}
\end{center}
\end{figure}

\begin{figure}[h]
  \begin{center}
  \includegraphics[width=\columnwidth]{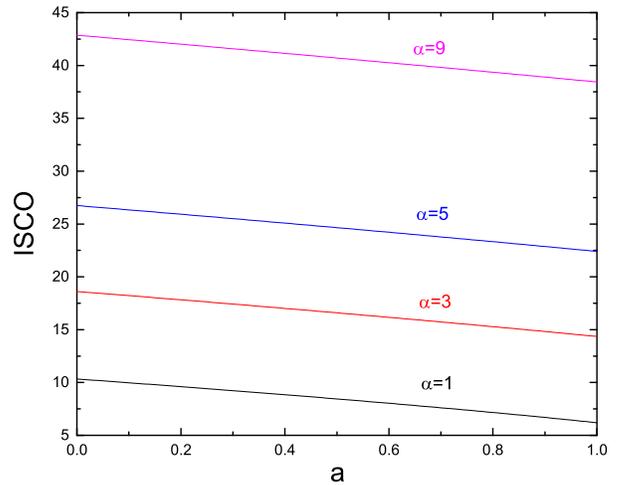}
  \caption{{\bf ISCO of large parameter $\alpha$}. The radius of ISCO decreases with increasing values of $a$ and increases as $\alpha$ increases.}
  \label{fig.5}
\end{center}
\end{figure}

In the case of Kerr black hole, radius of ISCO is depend on intrinsic angular momentum $a$ of black hole. However, the radius of ISCO of Kerr-MOG is depend on $a$ and parameter $\alpha$ which is related to gravitational constant $G$.

Fig.1 shows that the minimum point of the energy of test particles is a radius of ISCO of Kerr-MOG black hole when $\alpha$ is $9$ and $a$ is $0.95$.
In case of Fig.2, parameter $\alpha$ is $0.5$ and $a$ is $0.95$.
In Ref. \cite{R11}, the event horizon $r_{+}$ given by

\begin{equation}
\begin{split}
r_{+}=G_{N}(1+\alpha)M\Bigg[1+\sqrt{1-\frac{a^{2}}{G_{N}^{2}(1+\alpha)^{2}M^{2}}-\frac{\alpha}{1+\alpha}}\Bigg].
\end{split}
\end{equation}

Fig.3 show that event horizons and ISCOs of Kerr-MOG black hole are proportional to parameter $\alpha$.
It shows that the distance between event horizon and ISCO of Kerr-MOG black hole gradually increases as parameter $\alpha$ increases. In the figure, the event horizons for $a=0.5$ and $a=0.95$ show little difference. In Ref.\cite{R12}, black hole shadow image depend on ISCO geometry. The radius of black hole shadow of Kerr-MOG $r_{shad}$ is calculated by $r_{shad}=22.68r_{s}$ when $\alpha =9$ and $a=0.95$, where $r_{s}$ is $2G_{N}(1+\alpha)M$. Comparing radius of ISCO with radius of black hole shadow of Kerr-MOG, we obtain $r_{shad}=11.729r_{ISCO}$.

In Fig.4, comparing Kerr with Kerr-MOG black hole, ISCOs of Kerr-MOG black holes are always greater than Kerr black hole.
ISCO of Kerr black hole converges $1$ at $a=1$, while ISCO of Kerr-MOG is not. Thus, for rapidly spinning black holes $(a>1)$, ISCO of Kerr-MOG can be greater than $1$. Moreover, ISCO of Kerr black hole is equal to event horizon when $a=1$, but ISCO of Kerr-MOG black hole is greater than its event horizon when $a=1$.

Fig.5 shows that the curves for ISCOs of high values of parameter $\alpha$.
$\alpha$ is larger, radius of ISCO get away at the center of black hole. Furthermore, curves for radius of ISCO of Kerr-MOG black hole become linear for intrinsic angular momentum $a$. ISCO is increased as parameter $\alpha$ increases.

\section{CONCLUSIONS}

In this paper, we investigated rotating black hole in modified gravity theory called Kerr-MOG black hole. In particular, we have calculated conserved quantities for a test particles of Kerr-MOG black holes by using to well-established method \cite{R18,R19} and numerically computed radius of ISCO in Kerr-MOG black hole. We also investigated the phenomena by parameter $\alpha$ by comparing with ISCO of Kerr black hole. For the Kerr-MOG black hole, radius of ISCO is depend on parameter $\alpha$ as well as intrinsic angular momentum $a$. ISCO of Kerr black hole for a test particle decreases as $a$ and ISCO of Kerr-MOG black hole increases as parameter $\alpha$ increases. Moreover, the distance of between the ISCO of Kerr-MOG black hole and event horizon also increase as parameter $\alpha$ increases. The ISCO of Kerr-MOG can help study comparing black hole shadow of kerr-MOG with astronomical observation of supermassive black holes like SgA* at the center of the Milky Way. it also can expect to radiate more around black hole as distance between ISCO and event horizon of black hole increases. In the near future, we are going to study for ISCO of Kerr-MOG of lightlike geodesics and combination of Non-Kerr and Kerr-MOG black hole. ISCO of Kerr-MOG black hole expect good use of comparison with experimental values of EHT.

\section{ACKNOWLEDGMENTS}

This work is supported by Research Fund of Soonchunhyang University.


\begin{references}
\bibitem{R1} F. Zwicky, \href{http://adsabs.harvard.edu/abs/1933AcHPh...6..110Z}{Helv. Phys. Acta {\bf6}, 110} (1933), \href{http://www.ymambrini.com/My_World/History_files/Zwicky.pdf}{english}.
\bibitem{R2} F. Zwicky, \href{http://adsabs.harvard.edu/abs/1937ApJ....86..217Z}{APJ, {\bf86}, 217} (1937).
\bibitem{R3} V. Rubin, W. K. J. Ford, \href{http://adsabs.harvard.edu/abs/1970ApJ...159..379R}{Astrophys. J. {\bf159}, 379} (1970).
\bibitem{R4} V. C. Rubin, N. Thonnard and W. K. J. Ford, \href{http://adsabs.harvard.edu/abs/1980ApJ...238..471R}{ Astrophys. J. {\bf 238}, 471} (1980).
\bibitem{R5} M. Milgrom, \href{http://adsabs.harvard.edu/abs/1983ApJ...270..365M}{Astrophys. J. {\bf270}, 365} (1983).
\bibitem{R6} Bhattacharjee, S. Chaudhury and S. Kundu, \href{http://iopscience.iop.org/article/10.1088/0004-637X/785/1/63/meta;jsessionid=BE28A3875F16DB58F304D5C3FAFECA4C.ip-10-40-1-105}{P. ApJ, {\bf63}, 785} (2014), \href{https://arxiv.org/abs/1310.2659}{arXiv:1310.2659}.
\bibitem{R7} J. W. Moffat, V.T. Toth, \href{http://iopscience.iop.org/article/10.1086/587926/meta}{ApJ, {\bf680}, 1158} (2008), \href{https://arxiv.org/abs/0708.1935}{arXiv:0708.1935}.
\bibitem{R8} J. W. Moffat, \href{http://iopscience.iop.org/article/10.1088/1475-7516/2006/03/004/meta}{JCAP, {\bf3}, 4} (2005), \href{https://arxiv.org/abs/gr-qc/0506021}{arXiv:gr-qc/0506021}.
\bibitem{R9} J. W. Moffat, \href{http://www.sciencedirect.com/science/article/pii/S0370269316306608}{Physics Letters B, {\bf763}, 427} (2016), \href{https://arxiv.org/abs/1603.05225}{arXiv:1603.05225}.
\bibitem{R10} J. W. Moffat, \href{https://arxiv.org/abs/1611.05382}{arXiv:1611.05382} (2016).
\bibitem{R11} J. W. Moffat, \href{https://link.springer.com/article/10.1140/epjc/s10052-015-3405-x}{Eur. Phys. J. {\bf C75}, 175} (2015), \href{https://arxiv.org/abs/1502.01677}{arXiv:1502.01677.}
\bibitem{R12} J. W. Moffat, \href{https://springerlink.altmetric.com/details/3373897}{Eur. Phys. J. {\bf C75} 130} (2015).
\bibitem{R13} A. E. Broderick, T. Johannsen, A. Loeb and D. Psaltis, \href{http://iopscience.iop.org/article/10.1088/0004-637X/784/1/7/meta}{Astrophys.J. 784, 7} (2014), \href{https://arxiv.org/abs/1311.5564}{arXiv:1311.5564}.
\bibitem{R14} S. Hussain and M. Jamil, \href{https://journals.aps.org/prd/abstract/10.1103/PhysRevD.92.043008}{Phys. Rev. D {\bf92}, 043008} (2015), \href{https://arxiv.org/abs/1508.02123}{arXiv:1508.02123}.
\bibitem{R15} M. A. Abramowicz and P. C. Fragile, \href{https://link.springer.com/article/10.12942/lrr-2013-1}{Living Rev. Relativity, 16, 1} (2013).
\bibitem{R16} T. Johannsen, \href{http://iopscience.iop.org/article/10.1088/0264-9381/33/12/124001/meta}{Class. Quant. Grav. {\bf33}, 124001} (2016), \href{https://arxiv.org/abs/1602.07694}{arXiv:1602.07694}.
\bibitem{R17} P. I. Jefremov, O. Y. Tsupko and G. S. Bisnovatyi-Kogan, \href{https://journals.aps.org/prd/abstract/10.1103/PhysRevD.91.124030}{Phys.Rev. D {\bf91}, 124030} (2015), \href{https://arxiv.org/abs/1503.07060}{arXiv:1503.07060}.
\bibitem{R18} S. Chandrasekhar, The Mathematical Theory of Black Holes(Oxford, England, 1992).
\bibitem{R19} C. Chakraborty, \href{https://doi.org/10.1140/epjc/s10052-014-2759-9}{Eur. Phys. J. {\bf C74}, 2759} (2014), \href{https://arxiv.org/abs/1307.4698}{	arXiv:1307.4698}.
\bibitem{R20} C. W. Misner, K. S. Thorne, J. A. Wheeler, Gravitation (W. H. Freeman, San Fransico, 1970).
\bibitem{R21} J. M. Bardeen, W. H. Press, and S. A. Teukolsky, \href{http://adsabs.harvard.edu/full/1972ApJ...178..347B}{Astrophys. J. 178, 347} (1972).
\bibitem{R22} D. C. Wilkins, \href{https://journals.aps.org/prd/abstract/10.1103/PhysRevD.5.814}{Phys. Rev. {\bf D5}, 814} (1972).
\bibitem{R23} D. Pugliese, H. Quevedo, and R. Ruffini, \href{https://journals.aps.org/prd/abstract/10.1103/PhysRevD.84.044030}{Phys. Rev. D {\bf 84}, 044030} (2011), \href{https://arxiv.org/abs/1105.2959}{arXiv:1105.2959}.
\bibitem{R24} A. W\"unsch, T. M\"uller, D. Weiskopf, and G. Wunner, \href{https://journals.aps.org/prd/abstract/10.1103/PhysRevD.87.024007}{Phys. Rev. D {\bf 87}, 024007} (2013), \href{https://arxiv.org/abs/1301.7560}{arXiv:1301.7560}
\bibitem{R25} D. Ayzenberg, N Yunes, \href{https://journals.aps.org/prd/abstract/10.1103/PhysRevD.90.044066}{Phys. Rev. D {\bf 90}, 044066} (2014), \href{https://arxiv.org/abs/1405.2133}{arXiv:1405.2133}.
\bibitem{R26} D. Pérez, G.E. Romero, S.E. Perez Bergliaffa, \href{http://www.aanda.org/articles/aa/pdf/2013/03/aa20378-12.pdf}{Astron. Astrophys. 551, A4} (2013), \href{https://arxiv.org/abs/1212.2640}{	arXiv:1212.2640}.
\bibitem{R27} T. Johannsen, and D. Psaltis, \href{https://journals.aps.org/prd/abstract/10.1103/PhysRevD.83.124015}{Phys. Rev. D. {\bf 83}, 124015} (2011), \href{https://arxiv.org/abs/1105.3191}{arXiv:1105.3191}.

\end{references}
\end{document}